\documentclass[doublecol]{epl2} 

\usepackage{amsmath}
\usepackage{graphicx}
\usepackage{dcolumn}
\usepackage{bm}
\usepackage{color}

\title{Dynamic properties of the one-dimensional Bose--Hubbard model}

\author{S.\ Ejima\inst{1} \and H.\ Fehske\inst{1} \and F.\ Gebhard\inst{2}}
\shortauthor{S.\ Ejima, H.\ Fehske and F.\ Gebhard}

\institute{\inst{1}Institut f\"ur Physik,
             Ernst-Moritz-Arndt-Universit\"at Greifswald,
             D-17489 Greifswald, Germany\\
\inst{2}Department of Physics, Philipps-Universit\"at Marburg, 
             D-35032 Marburg, 
             Germany}


\abstract{We use the density-matrix renormalization group method
to investigate ground-state and dynamic properties of the one-dimensional
Bose--Hubbard model, the effective model of ultracold bosonic atoms
in an optical lattice. For fixed maximum site occupancy $n_b=5$,
we calculate the phase boundaries between the Mott insulator 
and the `superfluid' phase for the lowest two Mott lobes. We extract
the Tomonaga--Luttinger parameter from the density-density correlation function
and determine accurately the critical interaction strength for the 
Mott transition. For both phases, we study the momentum distribution function
in the homogeneous system, and the particle distribution and
quasi-momentum distribution functions in a parabolic trap.
With our zero-temperature method we 
determine the photoemission spectra in the Mott insulator and in the
`superfluid' phase of the one-dimensional Bose--Hubbard model.
In the insulator, the Mott gap separates
the quasi-particle and quasi-hole dispersions.
In the `superfluid' phase the spectral weight
is concentrated around zero momentum.
\vspace*{1pt}}

\begin{document} 

\maketitle

\section{Introduction}
At very low temperatures, 
bosonic atoms which are loaded into an optical lattice
become superfluid for a shallow optical potential and Mott insulators
for a deep optical potential. The transition between both
phases has been observed experimentally~\cite{Greiner2002}; 
for a recent review, see~\cite{Zwerger2008}.
The Bose--Hubbard model provides a reasonable 
description of the experimental situation,
and its ground-state phase diagram in two and three dimensions
has been determined fairly accurately by perturbation
theory~\cite{Fisher89,Holthaus2008,Free09} and 
quantum Monte-Carlo (QMC) calculations~\cite{Prokofev07,Kato2008,Prokofev08}.

Bosons on a chain are also accessible experimentally~\cite{stoeferle2004}
so that it is interesting to study the
one-dimensional Bose--Hubbard model.
The physics in one dimension is rather peculiar. For example,
the state with the lowest kinetic energy is not macroscopically occupied
in the `superfluid'~\cite{Mora2003} but it is characterised 
by an algebraic divergence of the momentum distribution;
for a review, see~\cite{Giamarchi04}.
Moreover, the Mott gap is exponentially small in the Mott insulator
close to the phase transition. Therefore,
it is very difficult to determine the critical interaction strength
numerically. This problem also impairs the applicability
of strong-coupling perturbation theory.

In one dimension and at zero temperature, 
the density-matrix renormalisation group 
(DMRG) method~\cite{White92,km98,kwm00} permits the calculation
of ground-state properties with an excellent accuracy for large systems so that
the extrapolation to the thermodynamic limit can be performed reliably.
In this work, we use the density-density correlation function
to calculate the Tomonaga--Luttinger parameter from which
we determine the Mott transition accurately.
Moreover, we obtain the momentum distribution and the particle
distribution for bosons on a homogeneous chain and in the
presence of a harmonic trap.
Using the dynamical DMRG~\cite{j02}, we calculate
the single-particle spectral function at zero temperature
in the `superfluid' and the Mott insulating phases.

\section{Bose--Hubbard model}
The Hamilton operator
for the Bose--Hubbard model on a chain with an even number of sites~$L$ 
in a harmonic potential is defined by
\begin{eqnarray}
\hat{\cal H}&=&
 -t\sum_{j} \left( \hat{b}_j^{\dagger} \hat{b}_{j+1}^{\phantom{\dagger}}
          +\hat{b}_{j+1}^{\dagger} \hat{b}_{j}^{\phantom{\dagger}}\right)
 +\frac{U}{2}\sum_{j} \hat{n}_j(\hat{n}_j-1)
 \nonumber \\
 &&+V_c\sum_{j} (j-r_{\rm c})^2\, \hat{n}_j \; ,
  \label{hamil}
\end{eqnarray}
where $\hat{b}_j^{\dagger}$ and $\hat{b}_j^{\phantom{\dagger}}$ are 
the creation and annihilation operators for bosons on site~$j$,
$\hat{n}_j=\hat{b}_j^{\dagger} \hat{b}_j^{\phantom{\dagger}}$ 
is the boson number operator on site~$j$, $t$~is the tunnel amplitude between
neighbouring lattice sites, $U>0$~denotes the strength of the on-site
Coulomb repulsion, $V_c$ parameterises 
the curvature of the quadratic confining potential,
and $r_{\rm c}=(L+1)/2$ denotes the central position of the chain. 
In the following, we set $U=1$ as our energy unit, 
unless stated otherwise. 

\subsection{Constrained Bose--Hubbard model}
In general, the Bose--Hubbard model cannot be solved analytically.
In the low-density limit, the model reduces to 
the bose gas with $\delta$-potential interaction which was solved by
Lieb and Liniger~\cite{LL63}. The fact that three or more
bosons may occupy the same site forms the major obstacle on the way
to an exact solution. Since multiple-occupancies also pose 
technical problems in numerical approaches, 
the Bose--Hubbard model is usually approximated by the constraint
that there is a maximal number of bosons per site, $0 \leq n_b\leq {\cal N}-1$.
This constrained Bose--Hubbard model has ${\cal N}$ degrees of freedom
per site so that it can be written in terms of spin variables
with $S=({\cal N}-1)/2$. 
The case ${\cal N}=2$ is trivial because the hard-core Bose--Hubbard model
has no interaction term. It reduces to a model for free 
spinless fermions whose properties are known exactly~\cite{lsm61}.
The Bose--Hubbard model is recovered in the
limit ${\cal N}\to\infty$. In general, however,
the SU(${\cal N}$)-Bethe Ansatz equations 
do not solve the constrained Bose--Hubbard 
model~\cite{Frahm95,Frahm93}.

In our work, we study the restricted Bose--Hubbard model
with ${\cal N}=6$, i.e., $n_b\leq 5$. 
Our results are representative
for the original Bose--Hubbard model~(\ref{hamil}) 
because multiple lattice occupancies
are strongly suppressed in the parameter regions of interest to us,
$U/t>2$ and fillings $\rho=N/L< n_b$. 

\subsection{Numerical algorithm}
We adopt the DMRG method~\cite{White92} as 
our numerical tool for the calculation of
ground-state properties for constrained bose systems~\cite{km98,kwm00}.
For the spectral properties, we employ the dynamical DMRG (DDMRG)~\cite{j02}.

The considered lattices are large enough to permit reliable extrapolations
to the thermodynamic limit for the physical quantities of interest to us. 
We keep up to $m=2000$ density-matrix
eigenstates, so that the discarded weight is always smaller 
than $1 \times 10^{-10}$. 

We checked our algorithm for $n_b=1$ against the exact result~\cite{lsm61}.
The exact ground-state energy in the thermodynamical limit
and the extrapolated ground-state energy from DMRG
agree to four-digit accuracy.

\subsection{Ground-state phase diagram}
At integer filling $\rho=N/L$, the Bose--Hubbard model in one dimension 
describes a Mott transition between the `superfluid'
phase, characterised by a divergence of the momentum distribution
at momentum $k=0$~\cite{Mora2003}, 
and a Mott insulating phase, characterised by a finite
gap for single-particle excitations.
The latter is defined by the energy difference between the chemical
potentials for half band filling and one particle less than half filling,
\begin{eqnarray}
\Delta(L)&=& \mu^+(L)- \mu^-(L)\; , \nonumber \\
 \mu^+(L)&=&E_0(L,N+1)-E_0(L,N)\; , \\
 \mu^-(L)&=&E_0(L,N)-E_0(L,N-1)\; ,\nonumber
\end{eqnarray}
where $E_0(L,N)$ is the ground-state energy for $L$ sites
and $N$ particles. In the thermodynamical
limit, $N,L\to \infty$ and $\rho=N/L$ integer,
the gap is finite for the Mott insulator, 
$\Delta=\lim_{N,L\to\infty}\Delta(L)>0$,
so that the system becomes incompressible
when we go from the `superfluid' phase to the Mott insulating phase.

\begin{figure}[t]
 \begin{center}
  \includegraphics[scale=0.4]{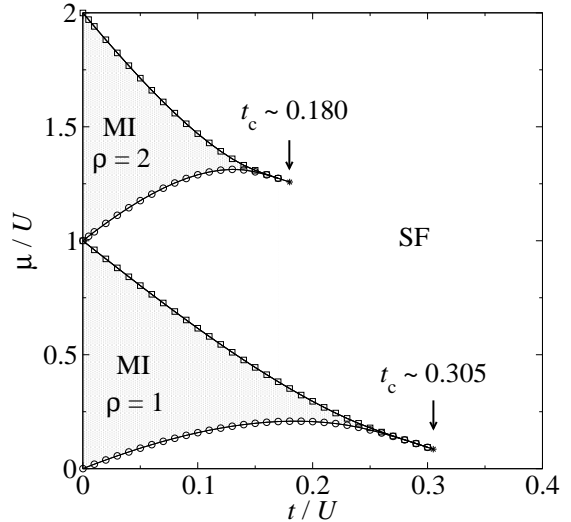}
 \end{center}
 \caption {Phase diagram of the one-dimensional
constrained Bose--Hubbard model ($n_b\leq 5$) from DMRG
with `superfluid' (SF) and Mott insulating (MI) regions.
The symbols confine the regions with a finite Mott gap, $\Delta>0$,
extrapolated from $\Delta(L)$ for $L\leq 128$.
The position of the Mott tips has been obtained from the 
Tomonaga--Luttinger parameter.\label{fig1}}
\end{figure}

The Mott transition lines in the $\mu$--$U$ ground-state phase diagram
have been previously 
determined by various analytical and numerical methods,
e.g., strong-coupling expansions~\cite{fm96,em99}, 
variational cluster approach~\cite{kd06}, 
QMC~\cite{bs92,kks96}, and DMRG~\cite{km98,kwm00}. 
In fig.~\ref{fig1} we show the phase diagram for the first 
Mott lobe ($\rho=1$) and the second Mott lobe ($\rho=2$) 
as obtained from our DMRG calculations with system sizes up to $L=128$. 

The overall shape of the Mott lobes agrees with previous results.
Here, we provide accurate data for the second Mott lobe, 
and the values for the critical
interaction strength for the first two Mott lobes
which we obtain from the Tomonaga--Luttinger parameter.
At the tip of each Mott lobe,
the model is in the universality class of the XY spin model so that
there is a Kosterlitz--Thouless phase transition with the
Tomonaga--Luttinger parameter $K_b=1/2$,
and the gap is exponentially small in the vicinity of
$(t/U)_c$. In contrast, SU(${\cal N}$)-Bethe Ansatz equations 
predict a discontinuity of the gap at the critical interaction
for ${\cal N}\geq 3$~\cite{Frahm95,Frahm93}.

\section{Ground-state properties}
\subsection{Tomonaga--Luttinger parameter and critical interactions for the
Mott transition}
The low-energy excitations of interacting bosons in
the superfluid phase are gapless linear excitations
(`phonons'). As in the case of fermionic systems
in one dimension~\cite{ssc02,egn05},
the Tomona\-ga--Luttinger parameter $K_b$ determines
the asymptotic behaviour of the correlation functions in the `superfluid' phase,
and various correlations functions have been used
to extract $K_b$~\cite{h81,g92,mfz08,sgci10}.
Here, we employ the density-density correlation function which is
defined by the ground-state expectation value
\begin{equation}
C(r)= \frac{1}{L}\sum_{\ell=1}^{L}
          \langle\hat{n}_{\ell+r}\hat{n}_{\ell}\rangle
           -\langle\hat{n}_{\ell+r}\rangle
            \langle\hat{n}_{\ell}\rangle \; .
\end{equation}
Asymptotically, it behaves like
\begin{equation}
 C(r\to \infty)\sim -\frac{1}{2K_b}\frac{1}{(\pi r)^{2}}
            +\frac{A\rho^2\cos(2\pi\rho r)}{(\rho r)^{2/K_b}}
            +\cdots \; .
\label{Cr-asymptotic}
\end{equation}
Thus, we can extract $K_b$ from the derivative
of its Fourier transformation,
\begin{equation}
\tilde{C}(q) = \sum_{r=1}^{L}e^{-{\rm i}qr}C(r) \; , \; 
0\leq q < 2\pi\; ,
\end{equation}
as $q=0$. In the thermodynamic limit one finds 
\begin{equation}
\frac{1}{2 \pi K_b}=\lim_{q\to 0}\frac{\tilde{C}(q)}{q} \; .
\label{eqn:krho}
\end{equation}
In order to treat finite systems in numerical calculations~\cite{egn05}, 
we translate~(\ref{eqn:krho}) into
\begin{equation}
 \frac{1}{2K_b(L)}=\lim_{L\to\infty}
                 \frac{L}{2}\tilde{C}\left(\frac{2\pi}{L}\right)\;,
\end{equation}
and extrapolate $K_b(L)$ to the thermodynamical limit.

In refs.~\cite{kwm00} and~\cite{zak08} 
the transition point has been also determined 
from the Luttinger parameter $K_b$. However, these authors estimated 
$K_b$ from the single-particle density matrix
\begin{equation}
 \Gamma(r)=\langle \hat{b}_r^{\dagger}\hat{b}_0^{\phantom{\dagger}}\rangle
          \sim r^{-K_b/2} \; \hbox{for $r \gg 1$}\; .
\label{gammakuehnerwhite}
\end{equation}
In their work, the extrapolation for the critical point $t_c$ depends on
the interval used for the fits to $\Gamma(r)$, see table~I
in ref.~\cite{kwm00}. When we derive the Luttinger parameter 
from eq.~(\ref{eqn:krho}) we can avoid this problem.

\begin{figure}[htbp]
 \begin{center}
  \includegraphics[scale=0.30,clip]{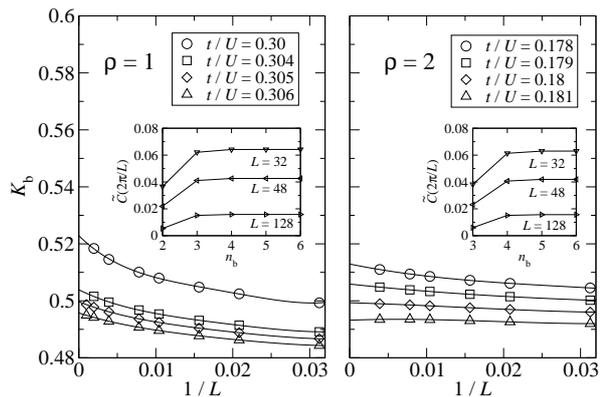}
 \end{center}
 \caption{Finite-size scaling 
for the Tomonaga--Luttinger parameter $K_b$ 
in the one-dimensional constrained Bose--Hubbard model ($n_b= 5$)
for the first ($\rho=1$) and second ($\rho=2$) Mott lobes, 
  using the DMRG with open boundary conditions. 
The lines are polynomial fits.\label{fig2}
The insets give the $n_b$-dependence 
of $\tilde{C}(2\pi/L)$ for various system sizes at $t/U=0.305$ 
(left panel) and $t/U=0.18$ (right panel).}
\end{figure}

As shown in fig.~\ref{fig2}, $K_b(L)$ can be reliably extrapolated 
to the thermodynamic limit using
polynomial functions in $1/L$. 
For $\rho=1$, we clearly have $K_b(t/U=0.3)>1/2$.
When we extrapolate our data for up to $L=1024$ lattice sites,
we find $K_b(t/U=0.304)>1/2$ but $K_b(t/U=0.306)<1/2$.
Therefore, we locate the transition point at
$t_c=0.305\pm 0.001$ for the first Mott lobe.
In the same way we find the transition point 
for the second Mott lobe at $t_c=0.180\pm0.001$
for the restricted Bose--Hubbard model with $n_b\leq 5$.
Using the same method, we have verified numerically 
that the values for the critical coupling $(t/U)_c$ are the same
for $n_b=4,6$ within our extrapolation uncertainty.

Note that $K_b(t<t_c)$ is not defined because we are in the
Mott insulating phase. However, $K_b(L)$ is finite and continuous over
the Kosterlitz--Thouless transition because 
the Mott gap is exponentially small near $t_c$. 
Nevertheless, our approach remains applicable
as has been shown for various 
fermionic models in refs.~\cite{egn05,satosato07}.

Previous groups located the Kosterlitz--Thouless transition for the
first Mott lobe at values consistent with ours.
In their DMRG work~\cite{kwm00}, K\"uhner {\it et al.} computed 
the Luttinger parameter using their DMRG algorithm on lattices
with up to $L=1024$ sites. From their fit to $\Gamma(r)$, 
eq.~(\ref{gammakuehnerwhite}),
they found $t_c=0.297\pm0.01$. Based on the same correlation function,
Zakrzewski and Delande~\cite{zak08} gave $t_c=0.2975\pm 0.005$ for the first
and $t_c=0.175\pm 0.002$ for the second Mott lobe for $n_b=6$. 
The determined $t_c$-values of such a kind significantly depend
on the interval of $r$, which is not the case within our approach. 
L\"auchli and Kollath~\cite{lk08} determined the critical point from 
the block entropy, by combining the recently developed quantum
information theory with the DMRG. 
Our result is within their region of the estimated values for $t_c$ 
(see fig.~2 in ref.~\cite{lk08}). 
In a combination of an exact diagonalisation study for systems 
with up to $L=12$ sites and a renormalisation group approach,
Kashurnikov and Svistunov found $t_c=0.304\pm0.002$~\cite{ks96}, and
their QMC calculations together with Kravasin gave
$t_c=0.300\pm0.005$~\cite{kks96}.
Another QMC calculation in combination with a 
renormalisation-group flow analysis of the finite-temperature data gave
$t_c=0.305(4)$~\cite{Lode08}, in perfect agreement with the result
of our zero-temperature study.

\begin{figure}[t]
 \begin{center}
  \includegraphics[scale=0.36,clip]{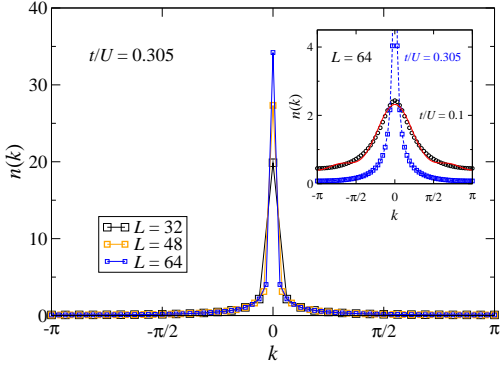}
 \end{center}
 \caption{(Colour online)
 Finite-size dependence of the momentum distribution function 
 $n(k)$ for $t=0.1$ (Mott insulator) and $t=0.305$ (`superfluid') 
in the one-dimensional Bose--Hubbard model
 using the DMRG with periodic boundary conditions. 
Inset: $n(k)$ for $L=64$. The solid line in the inset gives 
the result of strong-coupling theory to third order, 
eq.~(\protect\ref{nkstrong})~\protect\cite{Damski2006,Free09}.\label{fig3}}
\end{figure}

\subsection{Momentum distribution function}
Using the DMRG, the momentum distribution function
$n(k)$ can be calculated by taking the Fourier transformation of the 
single-particle density matrix 
\begin{equation}
n(k) =\frac{1}{L}\sum_{j,\ell=1}^{L}e^{{\rm i}k(j-\ell)}
        \langle \hat{b}^{\dagger}_{j} \hat{b}^{\phantom{\dagger}}_{\ell}
        \rangle\; ,
\label{eq:nk}
\end{equation}
where $k=2\pi m/L$ for $m=-L/2-1,\dots,L/2$ holds for periodic boundary
conditions~\cite{Kollath04}. 
Note that the momentum distribution function fulfils 
the sum rule $\sum_k n(k)=N$. In all cases of fig.~\ref{fig3} the 
numerical deviation
$\xi=|N-\sum_k n(k)|$ is always small, $\xi<1.0\times 10^{-3}$.

The difference between the superfluid phase and the 
Mott insulator is most markedly seen in the momentum distribution
$n(k)$ at momentum $k=0$: in the insulating phase, $n(k=0)$ remains finite
whereas it diverges as a function of system size in the superfluid phase,
as shown in fig.~\ref{fig3}. At $t/U=0.1$, $n(k=0)$ is almost
independent of system size, and the momentum distribution $n(k)$
is a smooth function of momentum~$k$.
Strong-coupling perturbation theory to third order~\cite{Damski2006,Free09}
predicts ($x=t/U$)
\begin{eqnarray}
n^{[3]}(k)&=&1+2C_1\cos(k) + 2C_2\cos(2k)+2C_3\cos(3k) \;,
\nonumber \\
C_1&=&4x-8x^3 \; , \; 
C_2= 18x^2\; , \; C_3=88x^3 \; .
\label{nkstrong}
\end{eqnarray}
Our numerical results for $t/U=0.1$ favourably compare with this expression,
see the inset of fig.~\ref{fig3}.

At $t/U=0.305$, above the critical point, $n(k=0)$ 
increases rapidly with system size.
In one spatial dimension there is no true superfluid 
with a macroscopic value for $n(k=0)$ in the thermodynamic 
limit~\cite{Mora2003}. Instead, we have from~(\ref{gammakuehnerwhite})
$n(|k|\to 0)\sim |k|^{-\nu}$, $\nu=1-K_b/2<1$.

\subsection{Local densities for the Bose--Hubbard model in a trap}
In the presence of the confining potential $V_c$ 
in the model~(\ref{hamil}), the density profile over the trap
is no longer homogeneous, see, e.g., ref.~\cite{brsrmdt02}.
For an open system, we define the quasi-momentum distribution
$\widetilde{n}(q)=
\langle \hat{b}^{\dagger}(q)\hat{b}(q)\rangle$
using the quasi-momentum states of particles in a box,
\begin{equation}
\hat{b}(q)=\sqrt{\frac{2}{L+1}}\sum_\ell\sin(q\ell)\hat{b}_{\ell}
\label{pseudomomenta}
\end{equation}
with $q=\pi n_q/(L+1)$ for integers $1\leq n_q\leq L$.

\begin{figure}[t]
 \begin{center}
  \includegraphics[scale=0.27]{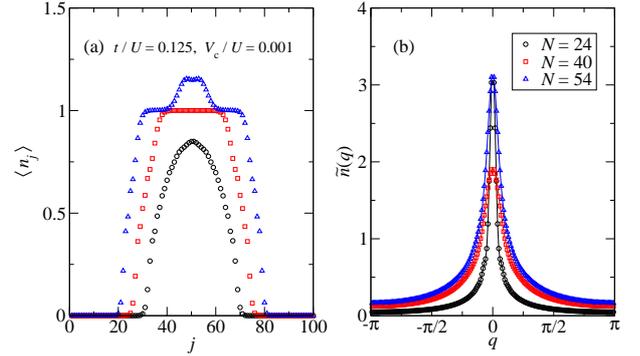}
 \end{center}
 \caption{(Colour online) Occupation probabilities 
in the one-dimensional constrained Bose--Hubbard model ($n_b\leq 5$)
in a parabolic trap potential of strength $V_c/U=0.001$.
We show the results for $N=24, 40, 54$ and 
$L=100$ ($\rho=0.24, 0.40, 0.54$)
for $t/U=0.125$ 
for (a) the local densities 
$\langle \hat{b}_j^{\dagger}\hat{b}_j^{\protect\phantom{\dagger}}\rangle$
and (b) the pseudo-momentum distribution $\widetilde{n}(q)$.
\label{figharmo}}
\end{figure}

As demonstrated by Batrouni {\it et al.}~\cite{brsrmdt02}
and Kollath {\it et al.}~\cite{Kollath04},
the potential confines the particles in the middle of the trap.
For small fillings, the local occupancies
display a bell-shaped distribution, where the maximum does not reach
the Mott plateau value, $\langle \hat{b}_j^{\dagger}
\hat{b}_j^{\phantom{\dagger}}\rangle(\rho=0.24)<1$.
The quasi-momentum distribution $\widetilde{n}(q)$ for this
superfluid in a trap shows a prominent peak at $k=0$.
For a larger filling, $\rho=0.40$, there exists
a Mott plateau, $\langle \hat{b}_j^{\dagger}
\hat{b}_j^{\phantom{\dagger}}\rangle(\rho=0.40)=1$ for $40<j<60$.
Recall that, for $t/U=0.125$, the homogeneous system at filling $\rho=1$
is a Mott insulator. Correspondingly, the peak in the pseudo-momentum
distribution at $k=0$ is smaller for $\rho=0.40$ than for $\rho=0.24$, see
fig.~\ref{figharmo}~(b). Finally, at filling $\rho=0.54$,
the confining potential and the bosons' tendency to cluster overcome
the repulsive potential in the middle of the trap
so that local occupancies larger
than unity are seen inside the trap. Correspondingly,
the peak intensity of the pseudo-momentum distribution at $\rho=0.54$
exceeds its value for $\rho=0.40$. 

\section{Photoemission spectra}
Single-particle excitations associated with the injection or emission of
a boson with wave vector $q$ and frequency $\omega$, 
$A^{+}(q,\omega)$ or $A^{-}(q,\omega)$, 
are described by the spectral functions
\begin{equation}
 A^{\pm}(q,\omega)=\sum_n|\langle\psi_n^{\pm}|\hat{b}^{\pm}(q)|\psi_0\rangle^2
  \delta(\omega\mp\omega^{\pm})\; ,
\label{Akw}
\end{equation}
where $\hat{b}^+(q)=\hat{b}^{\dagger}(q)$ and 
$\hat{b}^{-}(q)=\hat{b}(q)$ create/annihilate
particles with pseudo-momentum~$q$.
Moreover, $|\psi_0\rangle$ is the ground state of a $L$-site system in the 
$N$-particle sector while $|\psi_n^{\pm}\rangle$ denote the $n$th
excited states in the $(N\pm 1)$-particle sectors with excitation energies
$\omega_n^{\pm}=E_n^{\pm}-E_0$.

So far, very few data are available for the (inverse) photoemission spectra 
in the one-dimensional (constrained) Bose--Hubbard model. 
Analytical results include the variational cluster perturbation 
theory~\cite{kd06}, the random phase approximation~\cite{mt08}, and 
strong-coupling theory~\cite{Oosten2001}.
Pippan {\it et al.}~\cite{peh09} combined QMC
at low but finite temperatures with the maximum-entropy method 
to extract the spectral functions.

In the following we present the (inverse)-photoemission spectra at 
zero temperature using the numerically exact dynamical DMRG 
method~\cite{j02,jf07}.
We keep $m= 500$ states to obtain the ground state
in the first five DMRG sweeps and take $m=200$
states for the calculation of the various spectra from~(\ref{Akw})
by DDMRG. 
For a bosonic system the following sum-rules hold, 
\begin{eqnarray}
 \int_{-\infty}^{\infty}{\rm d}\omega 
\left(A^+(k,\omega)-A^-(k,\omega)\right)
&=& 1 \; , \label{sumrule} \\
\int_{-\infty}^0{\rm d}\omega 
\left(A^+(k,\omega)+A^-(k,\omega)\right) &=& n(k)\; .
\label{sumrule-nk} 
\end{eqnarray}
In our DDMRG calculations, both sum-rules are fulfilled with high precision.

\begin{figure}[t]
 \begin{center}
  \includegraphics[clip,scale=0.42]{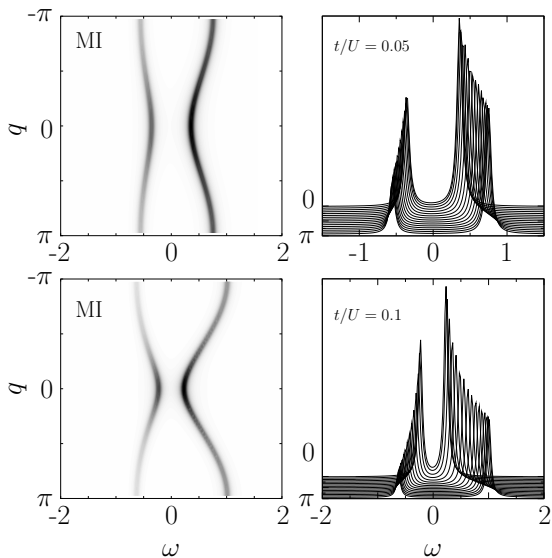}
 \end{center}
 \caption{Intensity (left panels) and line-shape (right panels) of the
 single-boson spectral functions $A(q,\omega)$ 
 in the Mott insulating (MI) phase
for $t/U=0.05$ (upper panels) and $t/U=0.1$ (lower panels) 
 with system size $L=64$ at filling $\rho=1$ 
using the DDMRG technique with open boundary conditions.
\label{fig5}}
\end{figure}

In fig.~\ref{fig5} we show the results for the Mott insulator with $\rho=1$.
The spectra $A(q,\omega)=A^+(q,\omega)+A^-(q,\omega)$ 
for fixed $q$ consist of
two Lorentzians of width $\eta=0.04$,
the size of the broadening introduced in the DDMRG procedure.
The quality of the fits suggests that the quasi-particle life-time
is very large in the Mott insulator. 

\begin{figure}[t]
 \begin{center}
  \includegraphics[clip,scale=0.35]{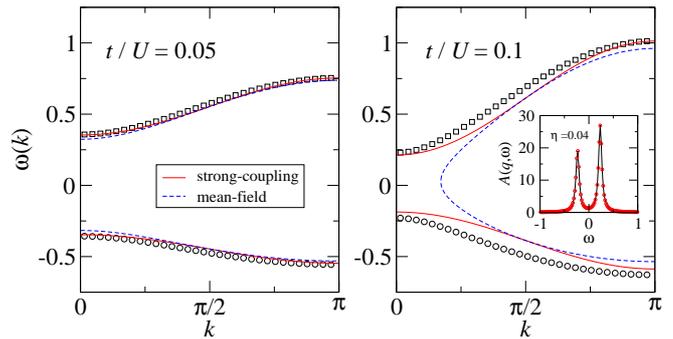}
 \end{center}
 \caption{(Colour online) Quasi-particle dispersions $\omega(k)$
in the Mott insulating phase at filling $\rho=1$ 
for $t/U=0.05$ and $t/U=0.1$
from Lorentz-fits to the spectral functions 
$A(q,\omega)$. 
For comparison, we also show the strong-coupling dispersions
for the propagation of a hole and a double occupancy,
$\omega_{\rm h,p}(k)$, and the mean-field result of 
ref.~\cite{Oosten2001}.\label{fig6}
Inset: $A(q,\omega)$ for $t/U=0.1$ at $q=\pi/33$ with $L=32$ (line)
and $q=2\pi/65$ with $L=64$ (circles) demonstrating the negligible system
size dependence.
}

\end{figure}
\begin{figure}[h]
 \begin{center}
  \includegraphics[scale=0.45,clip]{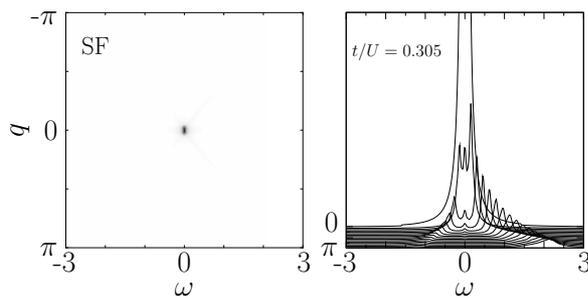}
 \end{center}
\caption{Intensity (left panels) and line-shape (right panels) of the
single-boson spectral functions 
$A(q,\omega)$  
for $t_c=0.305$ (`superfluid' phase)
with system size $L=64$ at filling $\rho=1$
using the DDMRG technique with open boundary conditions for a
broadening $\eta=0.04$\label{fig7}}
\end{figure}

In fig.~\ref{fig6} we show the quasi-particle dispersions for $t/U=0.05$ 
and $t/U=0.1$
which we extracted from the fits of the spectral functions to two Lorentz peaks
at $\omega=\omega^{\pm}(k)$.
For comparison, we include the mean-field result~\cite{Oosten2001}
and the strong-coupling result. 
For large interactions, each site is singly occupied in the
ground state. A hole excitation can propagate freely so
that the dispersion relation is given by $\omega_{\rm h}(k)=-\mu+2t\cos(k)$.
Likewise, a doubly occupied site can
also move freely through the system. Since either of the two bosons
of the doubly occupied site can tunnel to its neighbouring sites,
the dispersion relation is given by 
$\omega_{\rm p}(k)=U-\mu-4t\cos(k)$. These expressions for the
quasi-particle dispersions are exact to leading and first order 
in strong-coupling perturbation theory.

In fig.~\ref{fig7} we show the spectral functions in the `super\-fluid'
phase for $\rho=1$ close to the Mott transition, $t=0.305$.
The elementary excitations concentrate around $(k=0,\omega=0)$.
This confirms the formation of a `condensate', as also seen in
the momentum distribution, see fig.~\ref{fig3}. Moreover,
it shows that the low-energy excitations near $k=0$
indeed dominate the spectral functions. We used
this concept for the analysis of the ground-state correlation functions,
see eq.~(\ref{Cr-asymptotic}).
Note that deep inside of the Mott phase the system size
dependence of the spectral functions is insignificant, see
inset of fig.~\ref{fig6}.

\section{Conclusions}
In this work we have investigated the one-dimensional constrained
Bose--Hubbard model ($n_b\leq 5$) at zero temperature. 
Using the density-matrix renormalisation
group method we have obtained the Tomonaga--Luttinger parameter~$K_b$
from the density--density correlation function and determined
the critical couplings $(t/U)_c=0.305(1)$ for density $\rho=N/L=1$
and $(t/U)_c=0.180(1)$ for density $\rho=2$ which separate the `superfluid'
and Mott insulating phases. 

In the `superfluid' phase,
the momentum distribution diverges for small momenta,
$n(|k|\to 0)\sim |k|^{-\nu}\sim L^{\nu}$ ($\nu=1-K_b/2$), 
and the spectral function 
is finite only for small frequencies and momenta.
In the presence of a confining potential, we recover 
the Mott plateau in the particle density 
for filling $\rho=0.40$ and the wedding-cake structure
for filling $\rho=0.54$.

In the Mott insulator, the momentum distribution
is a continuous function. The spectral function
is well described in terms of free quasi-hole
and quasi-particle excitations which have a very long life-time
for strong correlations.
Their dispersion relation can be obtained 
from strong-coupling perturbation theory. 
A calculation of the quasi-particle bands
beyond first order remains to be done.

\acknowledgments
The authors would like to thank H. Frahm and T. Giamarchi 
for valuable discussions. SE and HF acknowledge funding by the 
DFG through grant SFB~652.

\end{document}